\documentstyle[aps,multicol,fleqn]{revtex}
\input{amssym.def}

\title{Coherent States for Transparent Potentials}
\author{Boris F. Samsonov}
\address{Quantum Field Theory Dpt., Tomsk State University,\\
36 Lenin Avenue, 634050 Tomsk, Russia \\
email: samsonov@phys.tsu.ru}

\begin{document}

\maketitle

\hfill quant-ph/9904090

\begin{abstract}
Darboux transformation operators that produce multisoliton
potentials are analyzed as operators acting in a Hilbert space.
Isometric correspondence between Hilbert spaces of states of a
free particle and a particle moving in a soliton potential is
established. It is shown that the Darboux transformation
operator is unbounded but closed and can not realize an
isometric mapping between Hilbert spaces. A quasispectral
representation of such an operator in terms of continuum bases
is obtained. Different types of coherent states
of a multisoliton potential are introduced. Measures that
realize the resolution of the
identity operator in terms of the projectors on the
coherent states vectors
are calculated. It is shown that when
these states are related with
free particle coherent states by
a bounded symmetry operator the
measure is defined by ordinary functions and in the case of a
semibounded symmetry operator
the measure is defined by a generalized function.
\end{abstract}

\begin{multicols}{2}

\section{INTRODUCTION}

The concept of coherent states (CS) is widely used in different
fields of physics and mathematics (see for example Refs.
\cite{Per} - \cite{KS}). In particularly, it plays an important
role in the
Berezin quantization scheme \cite{Ber}, in the analysis of
growth
of functions holomorphic in a complex domain \cite{Vourd}, in a
general theory of phase space quasiprobability distributions
\cite{BM}, and in a quantum state engineering \cite{Vo}.
It is necessary to note that in present no a unified definition
of such states exists in the literature and different authors
mean different things when speaking about them. Nevertheless, a
careful analysis (see for example Ref. \cite{Klaud}) shows that
almost all definitions
have some common points that can be taken as a general
definition. Following Klauder \cite{Klaud} I mean by coherent
states such states that satisfy the following defining
properties:
(1) CS are defined by vectors $\psi _z(x,t)$ which belong to a
Hilbert space $H$ of the states of a quantum system
with scalar product $\langle \cdot |\cdot \rangle $;
(2) The parameter $z$ takes continuous values from a domain
${\cal D}$ of an $n$-dimensional complex space;
(3) There exists a measure $\mu =\mu (z,\bar z)$ (the bar over
a symbol indicates complex conjugation) that realizes the
resolution of the identity operator $\Bbb I$ acting in $H$
in terms of the projectors on the vectors $\psi _z$
\begin{equation}
\int _{\cal D} d\mu |\psi _z\rangle \langle \psi _z|=
\Bbb I\,;
\label{Id}
\end{equation}
(4)CS have to prove the property of a temporal
stability. By temporal stability I mean that the vectors
$\psi _z(x,t)$ remain coherent at all times i.e. satisfy the
properties 1.-3. at all times.
To satisfy this condition I shall
assume that the functions $\psi _z(x,t)$ are solutions of the
Schr\"o\-din\-ger equation
\[(i\partial _t-h_0)\psi _z(x,t) =0 \]
where $h_0$ is the Hamiltonian of a given quantum system which
in general can depend on time. Operator $h_0$ is supposed to be
Hermitian in $H$ and to have a unique self-adjoint extension.
The Eq. (\ref{Id}) should be understood in a weak sense.
This means that it is equivalent to the following relation
\[
\int _{\cal D} d\mu \langle \psi _a|\psi _z\rangle
\langle \psi _z|\psi _b \rangle =
\langle \psi _a|\psi _b\rangle
\]
which should hold for all $\psi _{a,b}$ from a dense set in
$H$.

Transparent potentials have many remarkable properties.
For instance, a quantum particle prove no reflection in the
scattering process on such a potential. Another remarkable
property is that each level in the discrete spectrum
of such a potential occupies a
preassigned position, which is controlled by values of the
parameters the potential depend on. The
discrete spectrum levels may even be
situated in the middle of the continuous spectrum. In the
latter case we have {\it completely transparent potentials}
\cite{Stahlh}.
Transparent potentials find a more significant application in
soliton theory. There is a marvelous vast literature on
this subject. I cite here only a monograph \cite{MatvSal}.
Because of their remarkable properties transparent potentials
would find an application in pseudopotential theories. Recently
they have been used to describe relaxation processes in Fermi
liquid \cite{VVT}.

CS for transparent potentials are very far from being explored.
It may be explained by the fact that up to now no systematic
way is
known for their investigation.
No a clear algebraic structure related to these potentials is
known and therefore well known algebraic methods \cite{Per}
prove
to be a little suitable in this context. No simple ladder
operators for the discrete spectrum eigenfunctions of
transparent potentials are known as well and therefor we can
not use the approach of Ref. \cite{MM} for this purpose. An
approach based on the uncertainty relation \cite{NiSai0} is not
consistent with the property 3. mentioned above and therefor it
should be rejected.

A conjecture has been advanced recently \cite{BSjetp,NF} to use
Darboux
transformation operator approach for investigating the CS of
those system that is related by Darboux  transformation
with that for which the CS are known.

Let us have an exactly
soluble Hamiltonian $h_0=-\partial _x^2 +V_0(x,t)$ for which
the CS $\psi _z(x,t)$ are known and we want to obtain the CS
for another Hamiltonian $h_1=-\partial _x^2 +V_1(x,t)$
related with $h_0$ by the Darboux  transformation operator that
I shall denote by $L$. In general it should be a {\it
nonstationary} Darboux  transformation operator defined by the
following intertwining relation \cite{BSPL}
\[L(i\partial _t-h_0)=(i\partial _t-h_1)L\, .\]
If such an operator $L$ is known then solutions of the
transformed Schr\"o\-din\-ger   equation
(i.e. the Schr\"o\-din\-ger   equation    with the
Hamiltonian
$h_1$) can easily be obtained by the action of the operator $L$
on solutions of the initial Schr\"o\-din\-ger   equation
(i.e. the Schr\"o\-din\-ger   equation
with
the Hamiltonian $h_0$). It is clear that the functions
$\varphi _z(x,t) =L\psi _z(x,t)$ will satisfy all the
properties of the
CS enumerated above except may be for the property 3. One of
the main goal of this paper is to prove that in the case of
soliton potentials this property is fulfilled.
I would like to mention that this approach has been
successfully
applied to study CS of anharmonic oscillator Hamiltonians with
equidistant and quasiequidistant spectra \cite{BSjetp} and CS
of
nonstationary soliton potentials \cite{jetp98E} that are
related with
soliton solutions of the Kadomtsev-Petviashvili equation. With
the help of this approach a classical counterpart of the
Darboux
transformation has been formulated and shown that at classical
level this transformation leads to a distortion of a phase
space \cite{SamJMP}. CS of a one-soliton potential have been
investigated as well and their supercoherent structure has been
revealed \cite{NF}. In this paper I generalize these results to
a multisoliton case.

The paper is organized as follows.
In the Section 2 I give well-known results for CS of the free
particle in preparation for their application in the following
sections. In the Section 3 the Darboux
transformation operator from the solutions of the
Schr\"o\-din\-ger   equation
with zero potential to the solutions of the same equation with
solitons potential is analyzed as an operator acting in the
Hilbert space of the states of a free particle. It is shown
that it
can not realize a mapping of Hilbert spaces since it is not
defined in the whole Hilbert space and can not be extended to
the whole Hilbert space. Isomeric operators expressed in terms
of continuous bases similar to these previously proposed by
L.D. ~Faddeev \cite{Fadd} and
analyzed by D.L. ~Pursey \cite{Pur} for the case of purely
discrete basis sets are introduced. These operators realize
a polar decomposition of Darboux  transformation operators. A
quazispectral representation of the Darboux  transformation
operator and
its inverse in terms of continuous bases are obtained. In the
Section 4. different systems of CS are introduced for soliton
potentials. It is established that the resolution of the
identity
operator exists in every case. Explicit expressions for
measures that realize this equality are found.
A brief conclusion brings a paper to a close.

\section{FREE PARTICLE COHERENT STATES}

In this section I give a brief overview of well-known
properties of the Hilbert space of states a free particle
(see Ref. \cite{Mil} and references therein) and corresponding
CS \cite{MM} we need for subsequent analysis.

Annihilation $a$ and creation $a^+$ operators
\[a=(i-t)\partial _x+ix/2,\quad  a^+=(i+t)\partial _x-ix/2\]
form the Heisenberg-Weil subalgebra of the six-dimensional
Schr\"o\-din\-ger
algebra which is a symmetry algebra of the equation    with
zero potential. Solutions of the free particle
Schr\"o\-din\-ger   equation
which are square integrable over full real axis
$\Bbb R =(-\infty , +\infty )$ with respect to the Lebesgue
measure are the eigenstates of the symmetry operator
$K_0=aa^++a^+a$, $K_0\psi _n(x,t) =(2n+1)\psi _n(x,t) $. Their
coordinate
representation is as follows
$$
\begin{array}{rl}
\psi _n(x,t) =
&
 (-i)^n(n!2^n\sqrt {2\pi })^{-1/2}
(1+it)^{-1/2} \\
\times &
\exp \!
\left[-in\arctan t+{\textstyle \frac {y^2}2}(it-1)\right]
H_n\left( y \right)\, ,\\
y=
&
{\displaystyle \frac{x}{\sqrt{2+2t^2}}}\, .\vphantom{\left)
{\displaystyle \frac AB} \right)}
\end{array}
$$
Operators $a$ and $a^+$ are the ladder operators for the basis
functions $\psi _n$:
$a\psi _n=\sqrt n\psi _{n-1}$,
$a^+\psi _n=\sqrt{n+1}\psi _{n+1}$, and $a\psi _0=0$.

By ${\cal L}_0 $ I denote the lineal of the functions $\psi_n
$,
$n=0,1,\ldots $ which is the space of all finite linear
combinations of the functions $\psi_n $ with the coefficients
from the field $\Bbb C$. The operators $a$ and $a^+$ being
linear are defined for all elements from ${\cal L}_0 $ and
${\cal L}_0 $ is
invariant with respect to the action of these operators.
Since the momentum operator $p_x=-i\partial _x$ and the initial
Hamiltonian $h_0$ are expressed in terms of $a$ and $a^+$:
$p_x=-(a+a^+)/2$, $h_0=p_x^2$, these operators are defined in
${\cal L}_0 $ and map this space into itself.

The Hilbert space of the states of the free particle, $H$, is
defined as a closure of the lineal ${\cal L}_0 $ with respect
to the
measure generated by the scalar product
$\langle \psi _a|\psi _b\rangle $, $\psi _{a,b}\in {\cal L}_0
$, which
is defined by the Lebesgue integral.
The functions $\psi _n$ form an orthonormal basis in $H$,
$\langle \psi _n|\psi _k\rangle =\delta _{nk}$.
It is well-known \cite{Smirn,RS}
that the operators $p_x$ and $h_0$ initially defined on ${\cal
L}_0 $
have unique self-adjoint extensions and consequently they
are essentially self-adjoint in $H$.
The spectrum of $h_0$ and $p_x$ is purely continuous. They have
common eigenfunctions $\psi _p=\psi _p(x,t)$:
$p_x\psi _p=p\psi _p$, $h_0\psi _p=p^2\psi _p$, $p\in \Bbb R$,
which do not belong to $H$ but belong to a more wide space
$H_-$ of the linear functionals over $H_+$, $H_+\subset
H\subset H_-$
(so called Gelfand triplet). We can choose the Hilbert-Schmidt
equipment of the space $H$ by letting $H_+=K_0^{-1}H$ since
$K_0^{-1}$ is a Hilbert-Schmidt operator. We refer a reader
to Refs. \cite{GSh,GV,BerezShub} for more details on the nested
Hilbert space.
The coordinate representation of the functions $\psi _p(x,t)$
is
well-known and I omit it here.

 The functions $\psi _p$ form an orthonormal and complete (in
the sense of generalized functions) basis in $H$,
$\langle \psi _p|\psi _q\rangle =\delta (p-q)$. The
completeness condition is expressed symbolically in terms of
the projectors onto these functions as follows
\begin{equation}
\int dp|\psi _p\rangle \langle \psi _p |=\Bbb I\, .
\label{PI}
\end{equation}
I do not indicate the limits of integration in the integrals
along the whole real axis. This equality should be understood
in a weak sense. This means that it is equivalent to
\[
\int dp\langle \psi _j|\psi _p\rangle
\langle \psi _p |\psi _k\rangle  =\delta _{jk}\, , \quad
j,k=0,1,\ldots
\]
where $\psi _k$, $k=0,1,\ldots $ are orthonormal basis
functions in the space $H$.

The free particle CS may be obtained by applying a displacement
operator in the Heisenberg-Weil group to the vacuum vector
$\psi _0$:
\[
\psi _z(x,t) =\exp (za^+-\bar z a)\psi _0(x,t)\,, \quad z\in
\Bbb C\,.
\]
These vectors are also the eigenvectors of the annihilation
operator $a\psi _z =z\psi _z$. The vectors $\psi _z\in H$
belong to a more wide set then ${\cal L}_0 $. Their Fourier
decomposition in terms of the basis $\psi_n$  has the form
\begin{equation}
\begin{array}{rl}
\psi _z=
&
\Phi \sum \nolimits _n a_n z^n \psi _n\,, \label {psz}
\end{array}
\end{equation}
\vspace{-4ex}
\[
\begin{array}{rl}
\Phi =
&
\Phi (z,\bar z)=\exp (-z\bar z /2)\,, \\
a_n=
&
(n!)^{-1/2},\quad z\in \Bbb C\,.
\end{array}
\]
The vectors $\psi _z(x,t) $ satisfy all the properties
enumerated in
the Introduction. In particular, the measure
$d\mu =d\mu (z,\bar z)$ from the relation (\ref{Id}) is well-%
known: $d\mu =dxdy/\pi $, $z=x+iy$ and the domain of
integration ${\cal D}$ is the whole complex plane $\Bbb C$. In
what follows I will not indicate the domain of integration in
the integrals over the measures. Integration will be always
extended over the whole complex plane. Finally I give a
coordinate representation of the free particle CS
$$
\begin{array}{rl}
\psi _z(x,t) =&
\hspace{-4pt}
(2\pi )^{-1/4}(1+it)^{-1/2}\\
&
\hspace{-14pt}
\times \exp \left[ -{\textstyle \frac 14}(z+\bar z)^2+
{\displaystyle \frac{(x+2iz)^2(it-1)}{4(1+t^2)}}  \right] .
\end{array}
$$

I use the notation $x$ as the spatial coordinate and as the
real part of a complex number $z$. I hope that it will not cause
a confusion since these quantities will never appear in the
same formula.

\section{DARBOUX TRANSFORMATIONS AND ISOMETRIC OPERATORS}

In this section I give an analysis of Darboux transformation
operator $L$ as an operator defined in the Hilbert space $H$.
I would like to stress that this operator is unbounded and can
not be defined over the whole space $H$. It has a domain of
definition which is a subset of $H$ and will be specified.
Moreover, it domain of values does not coincide with $H$.
Therefor this operator can not realize shifting between Hilbert
spaces
contrary to published assertion \cite{Montem}.

To obtain $N$-soliton potential I use the Darboux
transformation
operator approach elaborated in details in Ref. \cite{MatvSal}.
The
action of this operator on a sufficiently smooth function is
defined by the formula
\[L\psi =W^{-1}(u_1,\ldots ,u_N) W(u_1,\ldots ,u_N,\psi )\]
where $W$ stands for the usual symbol of a Wronskian. In the
case when the initial potential $V_0$ does not depend on time,
the functions $u_k=u_k(x,t)$ being solutions of the initial
Schr\"odinger  equation    may be eigenfunctions of the initial
Hamiltonian as well
$h_0u_k=a _ku_k$ and in general are not supposed to satisfy
any boundary conditions. In this case the transformation
operator $L$ does not depend on time and transforms solutions
of the initial Schr\"o\-din\-ger   equation    onto solutions
of the
Schr\"o\-din\-ger   equation    with the potential
\[V_1=V_0-2\partial _x^2\log W(u_1,\ldots ,u_N)\]
which is independent on time. In this paper we need not to use
time dependent Darboux  transformation which was proposed by
V. ~Matveev and M. ~Salle (see Ref. \cite{MatvSal}) and
advanced by V. ~Bagrov and B. ~Samsonov \cite{Rev}.

To obtain an $N$-soliton potential we should take $V_0=0$ and
specify the transformation functions $u_k$ as follows
\cite{MatvSal}:
\[
\begin{array}{rl}
u_{2k-1}=
&
\cosh (a_{2k-1}x+b_{2k-1})\,,\\
u_{2k}=
&
\sinh (a_{2k}x+b_{2k})\,,\\
h_0u_k=
&
-a_k^2u_k\,, \ k=1,2,\ldots ,N\,,\\
&
a_1<a_2<\ldots <a_N.
\end{array}
\]
The time dependent phase factors are omitted from these
functions since they do not affect all the results.
In general the Wronsky determinant contains $N!$ summands. I
would like to stress that in a special case of soliton
potentials this determinant may be substantially simplified and
presented as a sum of $2^{N-1}$ hyperbolic cosines \cite{CSh}
\[
W(u_1,\ldots ,u_N)=2^{1-N}\!\!\!\sum_{(\varepsilon_1,\ldots
,\varepsilon_N)}^{2^{N-1}}
\!\!\!\varepsilon_2\varepsilon_4\cdots\varepsilon_{p}\]
\vspace{-3ex}
\[\times \prod_{j>i}^{N}(
\varepsilon_{j}a_j-
\varepsilon_{i}a_i)\cosh [\sum_{l=1}^N\varepsilon_l
(a_{l}x+b_l)]\,,
\]
where $\varepsilon_i=\pm 1$, the value of the subscript $p$ at
$\varepsilon _p$ should be taken equal to $N$ for even $N$
values and to $N-1$ for odd $N$ values, the summation runs over
all ordered and nonidentical sets
$(\varepsilon_1,\ldots ,\varepsilon_N)$
(the sets $(\varepsilon_1, \ldots ,\varepsilon_N)$ and
$(-\varepsilon_1, \ldots ,-\varepsilon_N)$ are declared to be
identical).

It can be shown \cite{MatvSal} that the potential so obtained
is
regular and bounded from below. This implies that the
Hamiltonian $h_1=-\partial _x^2+V_1$ is essentially self-%
adjoint in $H$. It has a mixed spectrum. The position of the
discrete spectrum levels is defined by the values of the
parameters $a_k$: $E_k=-a_k^2$. Corresponding eigenfunctions
have the form \cite{BSTMF}
\[
\begin{array}{rl}
\varphi _k=
&
N_kW^{(k)}(u_1,\ldots ,u_N)/W(u_1,\ldots ,u_N)\,,\\
N_k=&
({\textstyle \frac 12}a_k\prod_{j=1(j\ne k)}^N|a_k^2-
a_j^2|)^{1/2} \,,\\
h_1\varphi _k=
&
-a_k^2\varphi _k, \quad k=1,\ldots ,N
\end{array}
\]
where $W^{(k)}(u_1,\ldots ,u_N)$ is the Wronskian of the
functions
$ u_1,\ldots ,u_N $ except for the function $u_k$
and the factor $N_k$ is introduced to ensure the normalization
of the functions $\varphi _k$,
$\langle \varphi _k|\varphi _j\rangle =\delta _{kj}$,
$k,j=1,\ldots N$.
The continuous spectrum corresponds to the semiaxis $E>0$.
Continuous spectrum eigenfunctions, $\varphi _p=\varphi _p(x,t)
$ of the
Hamiltonian $h_1$ may be obtained with the aid of the operator
$L$: $\varphi _p= N_p^{-1}L\psi _p$ where the factor
$N_p^{-1}$ such that
\[N_p^2=(p^2+a_1^2)\ldots (p^2+a_N^2)\]
is introduced to ensure
the normalization of the functions $\varphi _p$:
$\langle \varphi _p|\varphi _q\rangle =\delta (p-q)$,
$h_1\varphi _p=p^2\varphi _p$.
The set $\left\{\varphi _j,\ j=1,\ldots ,N;\ \varphi _p,\ p\in
\Bbb R\right\}$ is complete in $H$.

Since the operator $L$ is linear, the relation
$L\psi _p=N_p\varphi _p$ defines the action of this operator
on every $\psi $ of the form
\begin{equation}
\psi (x,t)=\int C(p)\psi _p(x,t)dp
\label{cp}
\end{equation}
where $C(p)$ is a finite continuous function over $\Bbb R$. The
set of functions of the form (\ref{cp}) is a linear space that
I shall denote by ${\cal L}_{0p}$ and it is dense (in the sense
of generalized functions) in $H$. (More precisely it is dense
in $H_-$ since these are functionals.) Hence, the action of the
operator $L$ is defined for every element from ${\cal L}_{0p}$.
The image of the space  ${\cal L}_{0p}$, that I shall denote by
${\cal L}_{1p}$ consists of the functions
\[
\varphi (x,t)=\int C(p)N_p\varphi _p(x,t)dp\, .
\]

The Darboux  transformation operator $L$ together with its
Laplace
adjoint $L^+$ has remarkable factorization properties
\cite{BSTMF,AIS}
\begin{equation}
g_0=L^+L=(h_0+a_1^2)\ldots (h_0+a_N^2)\, ,
\label{fac0}
\end{equation}
\vspace{-3ex}
\begin{equation}
g_1=LL^+=(h_1+a_1^2)\ldots (h_1+a_N^2)\, .
\label{fac}
\end{equation}

The functions $\psi _p$ are eigenfunctions of $g_0$,
$g_0\psi _p=N_p^2\psi _p$. This imply that the functions
$\varphi _p$ are eigenfunctions of the operator $g_1$,
$g_1\varphi _p=N_p^2\varphi _p$.
The discrete spectrum eigefunctions of the operator $h_1$,
$\varphi _k$, $k=1,\ldots ,N$ belong to the kernel of the
operator $g_1$, $g_1\varphi _k=0$, $k=1,\ldots ,N$. This means
that the operator $g_1$ is nonnegative in $H$. Therefor,
consider the orthogonal decomposition of the space $H$:
$H=H_0\oplus H_1$ where $H_0$ is an $N$-dimensional space with
the basis $\varphi _k$, $k=1,\ldots ,N$. The functions $\varphi
_p$, $p\in \Bbb R$ form a basis (in the sense of generalized
functions) in $H_1$. In what follows I shall not consider the
space $H_0$ and restrict the consideration only by the
space $H_1$. The operators $h_1$ and $g_1$ being restricted to
this space have only a continuous spectrum and the operator
$g_1$ is strictly positive. I conserve the same notations for
these operators as operators acting in $H_1$
Taking into account the spectral decomposition of these
operators
\[h_1=\int dpp|\varphi _p\rangle \langle \varphi _p|\, ,\]
\vspace{-4ex}
\[g_1=\int dpN_p^2|\varphi _p\rangle \langle \varphi _p| \]
we can specify their domain of definitions. For the operator
$h_1$ it consists of all $\varphi \in H_1$ such that the
integral
\[ \|h_1\varphi \|^2=
\int dpp^2|\langle \varphi|\varphi _p\rangle |^2 \]
converges and for the operator $g_1$ we should demand the
convergence of the integral
\[ \|g_1\varphi \|^2=
\int dpN_p^4|\langle \varphi|\varphi _p\rangle |^2\, .\]

It is clear that the operator $g_1$ is defined on ${\cal
L}_{1p}$ and maps this space into itself. Using this fact and
the factorization property (\ref{fac}) we can define the action
of the operator $L^+$ onto
the functions $\varphi _p$,
$L^+\varphi _p=N_p^{-1}L^+L\psi _p=N_p\psi _p$, and extend this
operator by linearity on the whole space ${\cal L}_{1p}$.

It is not difficult to see that the following equality
\[\langle L\psi _p|\varphi _q\rangle =
\langle \psi _p|L^+\varphi _q\rangle \]
holds for all $\psi _p\in {\cal L}_{0p}$ and
$\varphi _q\in {\cal L}_{1p}$. Nevertheless, this fact does not
mean that $L^+$ is an operator conjugate with respect to the
scalar product to $L$ which domain of definition is ${\cal
L}_{0p}$. To find such an operator we have to specify correctly
its domain of definition.
I shall not look for this domain. Instead I shall give a closed
extension $\bar L$ of the operator $L$ and then find its
conjugate
$\bar L^+$.

Once we know the bases $\psi _p$ and $\varphi _p$ in $H$ and
$H_1$ respectively we can consider isometric operators
\[U=\int dp |\varphi _p\rangle \langle \psi _p| \,,\]
\vspace{-4ex}
\[U^{-1}=U^+=\int dp |\psi _p\rangle \langle \varphi _p| \, .\]
Similar operators have been introduced by L.D. Faddeev
\cite{Fadd}
and considered by L. Pursey \cite{Pur} for purely discrete
bases.
These operators are bounded and defined for all elements from
$H$ and $H_1$ respectively.

Consider now the following operators
\begin{equation}
\bar L=\int dp N_p|\varphi _p \rangle \langle \psi _p|\, ,
\label{Lb}
\end{equation}
\vspace{-4ex}
\begin{equation}
\bar L^+=\int dp N_p|\psi _p \rangle \langle \varphi _p|\,.
\label{Lbk}
\end{equation}
It is not difficult to specify their domains of definition. For
this purpose I use the spectral decompositions of the operator
$g_0$ and its square root
\[g_0=\int dpN_p^2 |\psi _p\rangle \langle \psi _p|\, ,\]
\vspace{-3ex}
\begin{equation}
g_0^{1/2}=\int dpN_p |\psi _p\rangle \langle \psi _p|\, .
\label {g012}
\end{equation}
It follows that
\[\|\bar L\psi \|^2=\|g_0^{1/2}\psi \|^2=
\int dp N_p^2|\langle \psi |\psi _p\rangle |^2\,.\]
This means that the domain of definition of $\bar L$ coincides
with that of $g_0^{1/2}$ and consists of all $\psi \in H$ such
that the integral in the right hand side of this equation
converges. The domain of definition of $\bar L^+$ coincides
with that of the operator $g_1^{1/2}$. It is worthwhile to
mention that these domains may be described in terms of
conditions imposed on functions  that are comprised in these
domains (see for example \cite{KosSarg}) since $h_0$ and
$h_1$ are
operators bounded from below and essentially self-adjoint.

It is clear from the formulae (\ref{Lb}) and (\ref{Lbk})
 that the operator $\bar L^+$
is conjugate to $\bar L$ with respect to the scalar product and
their domains of definition are well specified. Moreover,
$\bar L^{++}=\bar L$. This imply \cite{Smirn,RS} that the
operator
$\bar L$ is closed. The formulae (\ref{Lb}) and (\ref{Lbk})
give quasispectral representation of the closed operators
$\bar L$ and $\bar L^+$.

It follows from the formulae (\ref{Lb}) and (\ref{Lbk})
that $\bar L\psi _p=L\psi _p=N_p\psi _p$ and
$\bar L^+\varphi _p=L^+\varphi _p=N_p\varphi _p$. This means
that $\bar L$ is a closed extension of the operator $L$ and
$\bar L^+$ is a similar extension of the operator $L^+$
when the domains ${\cal L}_{0p}$ and ${\cal L}_{1p}$ are
taken as their initial domains of definitions.

From the spectral decomposition of the operators $g_0^{1/2}$
(\ref{g012}) and $g_1^{1/2}$,
\[g_1^{1/2}=\int dp N_p|\varphi _p\rangle
\langle \varphi _p| \,,\]
we obtain the following representations for $\bar L$ and $\bar
L^+$:
\[\bar L=Ug_0^{1/2}=g_1^{1/2}U \, ,\quad
\bar L^+=g_0^{1/2}U^+=U^+g_1^{1/2}\, .\]
Such representations are known as {\it polar decompositions} or
{\it canonical representations} of
closed operators (see for example Refs. \cite{RS,DS}).

The action of the operator $U$ on the basis $\psi _n$ gives an
orthonormal basis in $H_1$:
$\zeta _n=U\psi _n$,
$\langle \zeta _n|\zeta_k\rangle =\delta _{nk}$. The functions
$\varphi _n=g_1^{1/2}\zeta _n=\bar L\psi _n=L\psi _n$, hence,
form a basis in $H_1$ equivalent to an orthonormal (so called
Riesz basis, see for example Ref. \cite{GK}). The operator $U$
is
nonlocal and rather complicated. Therefor there is no way in
which simple explicit expressions can be derived for the
functions $\zeta _N$. The functions $\varphi _n(x,t) =L\psi
_n(x,t)$ are
much simpler but they are not orthogonal to each other:
$\langle \varphi _n|\varphi _k\rangle =S_{nk}$.
I shall denote by $S$ the infinite matrix with the entries
$S_{nk}$. The elements of this matrix can easily be expressed
in terms of the elements of another matrix $S^0(a)$ with the
entries $S_{nk}^0(a)=\langle \psi _n|h_0+a^2|\psi _k\rangle $:
\[S_{nk}=\left[ S^0(a_1)S^0(a_2)\ldots S^0(a_N) \right]_{nk}\]
where the use of the factorization property (\ref{fac0}) has
been made. Taking into account that $h_0$ is expressed in terms
of the ladder operators $a$ and $a^+$ for the basis functions
$\psi _n$,
$h_0={\textstyle \frac 14} (a+a^+)^2$, we derive the nonzero
elements of
the matrix $S^0(a)$: $S_{nn}^0(a)=n/2+1/4+a^2$,
$S_{nn+2}^0(a)={\textstyle \frac 14}\sqrt {(n+1)(n+2)}$. All
the other
matrix elements are zero. We see, hence, that the number of
nonzero elements in each row and column of the matrix $S$ is
finite.

Consider now bounded operators
\[M=\int dp N_p^{-1}|\varphi _p\rangle \langle \psi _p| \,,\]
\vspace{-3ex}
\[M^+=\int dp N_p^{-1}|\psi _p \rangle \langle \varphi _p|\]
defined in $H$ and $H_1$ respectively. It is not difficult to
see that $M\bar L^+=\Bbb I$ is the unit operator in $H_1$ and
$M^+\bar L=\Bbb I$ is the unit operator in $H$. Using the
spectral resolutions of the operators $g_0^{-1/2}$ and $g_1^{-
1/2}$
\[g_0^{-1/2}=
\int dpN_p^{-1}|\psi _p\rangle \langle \psi _p| \,,\]
\vspace{-3ex}
\[g_1^{-1/2}=
\int dpN_p^{-1}|\varphi _p\rangle \langle \varphi _p|\]
we derive the polar decompositions of the operators $M$ and
$M^+$:
\[M=Ug_0^{-1/2}=g_1^{-1/2}U\,, \]
\vspace{-3ex}
\[M^+=g_0^{-1/2}U^+=U^+g_1^{-1/2}\,.\]
It is easily seen that these operators factorise the operators
inverse to $g_0$ and $g_1$: $M^+M=g_0^{-1}$, $MM^+=g_1^{-1}$.

The functions $\eta _n=g_1^{-1/2}\zeta _n=M\psi _n$ form
another basis in $H_1$ equivalent to an orthonormal. This basis
is
biorthogonal to $\varphi _n$,
$\langle \varphi _n|\eta _k\rangle =\delta _{nk}$.
It follows the representation for the elements $S_{nk}^{-1}$
$$
\begin{array}{rl}
S_{nk}^{-1}= &
\langle \eta _n|\eta _k \rangle =
\langle \psi _n|g_0^{-1}|\psi _k \rangle \\
=  &
{\displaystyle \int }dpN_p^{-2}\langle \psi _n|\psi _p\rangle
\langle \psi _p|\psi _k\rangle
\end{array}
$$

As a final remark of this section I would like to notice the
following.
The space $H_1$ can be obtained as a closure of the lineal
${\cal L}_1$ of all finite linear combinations of the functions
$\varphi _n=L\psi _n$ with respect to the norm generated by the
scalar product which is a restriction of the given scalar
product in $H$ to the lineal  ${\cal L}_1$. The set of
functions of the form $\varphi =\bar L\psi $ when $\psi $ run
through the whole domain of definition of the operator $\bar L$
(i.e. the domain $D_{\sqrt {g_0}}$ of definition of the
operator $\sqrt {g_0}$) can not give the whole space $H_1$.
Nevertheless, if we define a new scalar product in
 ${\cal L}_1$,
$\langle \varphi _a|\varphi _b\rangle _1\equiv
\langle L\psi _a|L\psi _b\rangle =
\langle \psi _a|g_0|\psi _b\rangle $,
$\psi _{a,b}\in {\cal L}_0$, $\varphi _{a,b}\in {\cal L}_1$
then the closure of  ${\cal L}_1$ with respect to the norm
generated by this scalar product will coincide with the set
$\varphi =\bar L\psi $, $\psi \in D_{\sqrt {g_0}}$. This space
is embedded in $H_1$.

\section{COHERENT STATES OF SOLITON POTENTIALS}

The operator $g_0$ is a symmetry operator for the
Schr\"o\-din\-ger   equation.
Therefor it commutes with the Schr\"o\-din\-ger   operator
$i\partial _t-h_0$
when applied to the solutions of the
Schr\"o\-din\-ger  equation. It follows that
the operator $U=\bar Lg_0^{-1/2}$ is an intertwining operator
for the Schr\"o\-din\-ger   operators $U(i\partial _t-
h_0)=(i\partial _t-h_1)U$
and therefor it is a transformation operator. Hence, being
applied to a solution of the initial
Schr\"o\-din\-ger  equation    (in our case the
free particle Schr\"o\-din\-ger   equation) it gives a solution
of the transformed
equation (in our case the Schr\"o\-din\-ger   equation    with
the $N$-soliton
potential). The functions $\zeta _n=U\psi _n $ and $\zeta
_z=U\psi _z$ are then solutions of the
Schr\"o\-din\-ger equation    with soliton
potential. The Fourier decomposition of the function $\zeta _z$
in terms of the basis $\left\{\zeta _n\right\}$ has the same
form as that of the function $\psi _z$ in terms of
$\left\{\psi _n\right\}$
\[\zeta _z=\Phi \sum\nolimits _n a_n\zeta _n\, .\]

The vectors $\zeta _z$, $z\in \Bbb C$ satisfy all the
conditions formulated for CS in the Introduction because of the
isometric nature of the operator $U$. The resolution of the
identity operator (\ref{Id}) in the space $H_1$ in terms of the
projectors on $\zeta _z$ takes place with the same measure
$d\mu =dxdy/\pi $, $z=x+iy$. One of the deficiencies of these
coherent states is that no a simple explicit expression for
such vectors exists. This deficiency may be cured by acting to
them by a symmetry operator for the
Schr\"o\-din\-ger equation    with soliton potential.

Consider the vectors
\[\varphi _z =g_1^{1/2}\zeta _z=\bar L\psi _z=
\Phi \sum \nolimits _na_n\varphi _n\, .\]
It is not difficult to see that the value
$\langle \psi _z|g_0|\psi _z\rangle $ is finite. This means
that $\psi _z $ belong to the domain of definition of the
operator $\bar L$ and the above equality has a sense. Moreover,
these functions are sufficiently smooth and we can apply to
them directly the differential operator $L$. Thus we obtain a
coordinate representation of $\varphi _z$. For instance, in the
case of the one-soliton potential this representation reads
$$
\begin{array}{rl}
\varphi _z(x,t)=  &
-{\textstyle \frac 12} (2\pi )^{-1/4}(1+it)^{-3/2} \\
\times  &
[x+2iz+2a(1+it)\tanh (ax)] \\
\times  &
\exp \left[ -{\displaystyle \frac {(x+2iz)^2}{4+4it}}-
{\textstyle \frac 14} (z+\bar z)^2
\right].
\end{array}
$$
We see thus that these functions are much simpler then $\zeta
_z$ and may be analyzed without difficulties. For example it is
easily seen that \cite{NF}
$|\varphi _z(x,t) |^2=|\varphi _z(-x,-t)|^2$. This property
reflects a
transparent nature of the one-soliton potential.

Another system of states may be obtained with the help of the
transformation operator $M$. Consider the vectors
\[
\eta _z=g_1^{-1/2}\zeta _z=M\psi _z=\Phi \sum\nolimits _na_n
\eta _n\, .
\]
The operator $M$ being inverse to $L$ has an integral nature.
For the case of the one-soliton potential the integration may
be carried out analytically \cite{NF}. This yields
$$
\begin{array}{rl}
\eta _z(x,t)= &
{\textstyle \frac {-i}{4}}\sqrt \pi (2\pi )^{-1/4}{\rm
sech}(ax)\\
\times  &
\exp \! \left[ -{\textstyle \frac 14}(z+\bar z)^2+a^2(1+it)
\right] \\
\times  &
\left[ \exp (2iaz){\rm erfc}\!
\left( a\sqrt {1+it}+{\displaystyle \frac {x/2+iz}{\sqrt
{1+it}}} \right)
\right. \\
- &
\left. \exp (-2iaz){\rm erfc}\!
\left( a\sqrt {1+it}-{\displaystyle \frac {x/2+iz}{\sqrt
{1+it}}} \right)
\right]\! .
\end{array}
$$
Where the parameter $b$ is taken to be zero.

It is worthwhile to mention that all the states
$\psi _z(x,t)$, $\varphi _z(x,t)$, $\eta _z(x,t)$, and $\zeta
_z(x,t)$
can not represent nonspreading in time wave packets.
Nevertheless, we can interpret them as coherent states since
they satisfy all the properties of such states enumerated in
the Introduction. I shall show now that for the vectors 
$\varphi _z$
and $\eta _z$ there exist measures
$\mu _\varphi =\mu _\varphi (z,\bar z)$ and
$\mu _\eta =\mu _\eta (z,\bar z)$ that realize the resolution
of the identity operator in $H_1$ in terms of the projectors on
these vectors.

First consider another continuous basis in $H_1$:
$\eta _p=N_pM\psi _p$,
$\langle \eta _p|\eta _q\rangle =\delta (p-q)$,
$p,q\in \Bbb R$. Since $\left\{\varphi _p\right\}$ and
$\left\{\eta _p\right\}$ are bases in $H_1$, the resolutions of
the identity operator of the type (\ref{Id}) in terms of the
vectors $\eta _z$ and $\varphi _z $ are equivalent to the
equations
$$
\begin{array}{c}
{\displaystyle\int }d\mu _\eta (z,\bar z)\langle \eta _p|\eta
_z\rangle
\langle \eta _z|\eta _q\rangle =\delta (p-q)\, ,\\
{\displaystyle \int }d\mu _\varphi (z,\bar z)\langle \varphi
_p|\varphi _z\rangle
\langle \varphi _z|\varphi _q\rangle =\delta (p-q)\, .
\end{array}
$$
Taking into account that the functions $\psi _p$ are the
eigenfunctions of $g_0$ and $g_0^{-1}$, $g_0\psi _p=N_p^2\psi
_p$, $g_0^{-1}\psi _p=N_p^{-2}\psi _p$ we arrive at equations
for the measures $\mu _\eta $ and $\mu _\varphi $
\hspace{-6em}\begin{equation}
(N_pN_q)^{-1}\int
d\mu _\eta \langle \psi _p|\psi _z\rangle
\langle \psi _z|\psi _q\rangle =\delta (p-q)\, ,
\label{mueta}
\end{equation}
\vspace{-3ex}
\hspace{-6em}\begin{equation}
N_pN_q \int
d\mu _\varphi \langle \psi _p|\psi _z\rangle
\langle \psi _z|\psi _q\rangle =\delta (p-q)\, .
\label{mufi}
\end{equation}
Note that the integrals involved in these equations are time-%
independent and hence can by calculated at $t=0$. Therefor in
what follows I let $t=0$ and look for the measures independent
on time.

The momentum representation of the CS $\psi _z$ is well-known
\[
\begin{array}{rl}
\langle \psi _p|\psi _z\rangle =
&
(2/\pi )^{1/4}\Phi \psi _p(z)\,,  \\
\psi _p(z)=
&
\exp (-p^2+2zp-z^2/2)\,,\  z=x+iy\,.
\end{array}
\]
Let us look for the measure $\mu _\eta $ in the form
$d\mu _\eta =\omega _\eta (x)dxdy$, $z=x+iy$. After performing
the integration with respect to $y$ in the Eq.
(\ref{mueta}) we arrive at an equation for $\omega _\eta (x)$
\[
(2\pi )^{1/2}\int dx \omega _\eta(x)F_p(x)=N_p^2\exp (2p^2)\,,
\]
\vspace{-3ex}
\[
F_p(x)=\exp (4px-2x^2)\,.
\]
The function $N_p^2$ is a polynomial in $p$ which is known. We
conclude then that $\omega _\eta (x)$ is a polynomial in $x$
whose coefficients are uniquely defined by the coefficients of
the polynomial $N_p^2$. For instance, for the one-soliton
potential we have
\[
\omega _\eta (x)=(x^2+a^2-1/4)/\pi\,.
\]
This proves that the states $\eta _z$ may be interpreted as CS.

We note that the states $\eta _z$ are defined with the help of
the bounded operator $g_0^{-1/2}$. This is the reason for which
the measure $\mu _\eta $ is expressed in terms of ordinary (non
generalized) functions. An other case takes place for the
states $\varphi _z $ which are defined by the semibounded
operator
$g_1^{1/2}$. I shall show now that the measure $\mu _\varphi $
is expressed in terms of generalized functions.

Let us look for the measure $\mu _\varphi $ in the form
$d\mu _\varphi =dyd\omega _\varphi (x)$. The integration in the
equation (\ref{mufi}) with respect to $y$ leads us to an
equation for the measure $d\omega _\varphi (x)$
\begin{equation}\label{Fp}
(2\pi )^{1/2}\int d\omega _\varphi (x)F_p(x)= N_p^{-2}
\exp (2p^2)\,.
\end{equation}
First we note that
$|F_p(x+iy)|\le \exp (-dx^2+by^2)$ where $2\le d\le b$.
This means that $F_p(x)$ belongs to a subspace of the space
$S_{1/2}^{1/2}$ of entire functions $F$ such that
$|F(x+iy)|\le \exp (-dx^2+by^2)$, $0\le d\le b$ \cite{GV}.
We look for $\omega _\varphi $ as a functional (i.e. a
generalized function) over $S_{1/2}^{1/2}$. (We will see that
really this is a functional over a subspace
${\stackrel{\circ}{{S}}}{}_{1/2}^{1/2}\subset
S_{1/2}^{1/2}$.)

As it is known \cite{GV} positive definite functionals (we look
for just such a functional) over $S_{1/2}^{1/2}$ are specified
by their Fourier transforms. Let $\tilde \omega _\varphi $ be
the Fourier transform of the measure $\omega _\varphi (x)$.
This means that an integration of a function
$F(x)\in S_{1/2}^{1/2}$ with respect to the measure
$\omega _\varphi (x)$ should be replaced by the integration
of the Fourier transform $\tilde F(t)$ of this function with
respect to the measure $\tilde \omega _\varphi $. In
particularly
\begin{equation}\label{Fpx}
\int d\omega _\varphi (x)F_p(x)=
\int d\tilde \omega _\varphi (t)\tilde F_p(t)
\end{equation}
where
$\tilde F_p(t)$ is the
Fourier image of the function $F_p(x)$ which in our case can
easily be found
\[\tilde F_p(t)=\sqrt {\pi /2}\exp (2p^2+ipt-t^2/8)\,.\]
As a result the Eq. (\ref{Fp}) yields the equation for
$\tilde \omega _\varphi (x)$
\[
\pi \int d\tilde \omega _\varphi (t)\exp (-t^2/8+ipt)=
N_p^{-2}\,.
\]
It is an easy exercise to see that $\tilde \omega _\varphi
(t)$
may be expressed in terms of elementary functions. For this
purpose we look for $\tilde \omega _\varphi (t)$ in the form
$d\tilde \omega _\varphi (t)=\rho _\varphi (t)dt$ and use the
following representation for the function $N_p^{-2}$:
\begin{equation}\label{Np}
\begin{array}{rl}
N_p^{-2}=
&
{\displaystyle \sum }_{k=1}^N {\displaystyle \frac {A_k}{\tau
+a
_k^2}}\,, \quad \tau=p^2\,,\\
A_k=
&
\left[(dN_p^2/d\tau )_{\tau =-a _k^2}\right]^{-1}.
\end{array}
\end{equation}
After some algebra we obtain a formula for $\rho _\varphi (t)$
\hspace{-6em}\begin{equation}\label{rofi}
\rho _\varphi (t)=
(2\pi )^{-1}\sum\nolimits _{k=1}^{N}\frac {A_k}{a _k}
\exp (t^2/8-a _k|t|)\,.
\end{equation}

Note that for the function $\rho _\varphi (t)$ of the form
(\ref{rofi}) there exist in $S_{1/2}^{1/2}$ such functions
$F(p)$ that
the integral in the right hand side of the Eq. (\ref{Fpx})
diverges. The convergence condition for this integral imposes a
restriction on the decrease of the integrand function $F(x)$
in the left hand side of the Eq. (\ref{Fpx}) as
$|x|\to \infty $. This function should satisfy an inequality
$|F(x)|\ge \exp (-2x^2-Ax)$ where $A$ is a nonnegative constant
own to every function $F(x)\in S_{1/2}^{1/2}$. I denote the set
of functions satisfying this condition by
${\stackrel{\circ }{{S}}}{}_{1/2}^{1/2}(\subset S_{1/2}^{1/2})$
which obviously is a linear space.

Thus, we have found the measure $\mu _\varphi $ in terms of the
generalized function
$\omega _\varphi (x)$ over the space
${\stackrel{\circ }{{S}}}{}_{1/2}^{1/2}$,
$d\mu_\varphi =dyd\omega _\varphi (x)$, $z=x+iy$ which is
defined by its Fourier transform $\tilde \omega _\varphi $.
The integrals with respect to this measure should be calculated
as follows
\[
\int d\mu _\varphi
\langle \varphi _a|\varphi _z\rangle
\langle \varphi _z|\varphi _b\rangle \equiv
\int dt \tilde \rho _\varphi (t)\tilde F_{ab}(t)
\]
where $ \tilde F_{ab}(t)$ is the Fourier transform of the
function
\[
F_{ab}(x)=\int dy
\langle \varphi _a|\varphi _z\rangle
\langle \varphi _z|\varphi _b\rangle \,,\
z=x+iy\,.\]

Finally I give comments on the calculation of the norms of
the functions $\eta _z$ and $\varphi _z$. The square of the
norm of $\eta _z$ may be calculated with the aid of the formula
(\ref{Np}) for the function $N_p^{-2}$ and the factorization
property of the operator $g_0^{-1}$ in terms of the operators
$M$ and $M^+$
\[
\langle \eta _z|\eta _z\rangle =
\langle \psi _z|g_0^{-1}|\psi _z\rangle =
\int dp N_p^{-2}|\langle \psi _z|\psi _p \rangle |^2\,.
\]
After some algebra we obtain
$$
\begin{array}{rl}
\langle \eta _z|\eta _z\rangle =
&
\sum \nolimits _{k=1}^NA_kF_k\,, \quad z=x+iy\,,\\
F_k=
&
{\displaystyle \frac {\sqrt {2\pi }}{a_k}}\exp [2(a_k ^2-x^2)]
 \vphantom {\left({\displaystyle \frac {\sqrt {2\pi
}}{a_k}}\right)}\\
\times
&
{\rm Re}\left[\exp (4ia_kx)
{\rm erfc}(a_k \sqrt 2+i\sqrt 2x)\right]\,.
\end{array}
$$

Similarly, the square of the norm of the function $\varphi _z$
coincides with the expectation value of the operator $g_0$ in
the state $\psi _z$. For instance, for the one-soliton
potential we obtain
$\langle \varphi _z|\varphi _z\rangle =
\langle \psi _z|g_0|\psi _z\rangle =
1/4+a^2+x^2$, $z=x+iy$.

\section{CONCLUSION}

A classical particle proves no reflection in the scattering
process on a potential well. For a quantum particle in general
this is not the case. Nevertheless, there exists a wide class
of potentials called transparent potentials for which the
scattering process of the quantum particle comes in some sense
about in a similar way that those of the classical particle
i.e. without reflection. In my opinion this mysterious
phenomena up to now has no any perspicuous explanation. From a
practical point of view the answer to this question is rather
important. If at quantum level we would be able to force a
signal to propagate without reflection we could decrease the
output of the emitted signal. All transparent potentials known
at present have a remarkable property. They are related with
zero potential (free particle) by Darboux transformations. Up
to recent times it was believed that such potentials have a
finite number of discrete spectrum levels. Nevertheless a
method based on an infinite chain of Darboux transformations
with the help of which one can create transparent potentials
with infinite number of discrete spectrum levels has been
proposed recently \cite{Shab}. To understand better the nature
of transparent potentials we should investigate them in all
details.

As it is well known the quantum theory gives a more detailed
description of the nature then the classical one. Therefor
different quantum systems may correspond to the same
classical system. 
Furthermore, the quantization
procedure is not unique (canonical quantization,
Berezin quantization, geometric quantization, etc.).
In this respect the following question is
of interest. What are common points between two classical
systems a quantization of which gives the quantum systems that
are related to each other by a Darboux transformation operator?
In particularly, what are common points between the classical
free particle and the particle that moves in a potential
quantization of which gives a transparent potential? The CS
approach make it possible to formulate clear steps in the
direction of obtaining an answer to this question. It permits
one to construct a classical mechanics counterpart of a given
quantum system and analyze properties of such a system. This
approach has been realized recently for the potential of the
form $x^2+gx^{-2}$ \cite{SamJMP}. It was established that at
classical level the Darboux transformation consists in a
distortion of a phase space of the classical system. Moreover,
this distortion is consistent with the transformation of the
Hamilton function in such a way that the equations of motion
remain unchanged.

Up to now no any approach for analysis of CS of transparent
potentials has been proposed. In this paper I show that the
Darboux  transformation operator approach is suitable for this
purpose. A next step in this direction would be an analysis of
the classical counterpart of the quantum system that moves in a
transparent potential.

\vspace{4ex}
\begin{center}
{\bf ACKNOWLEDGMENTS}
\end{center}

It is a pleasure to thank Dr. V.P. ~Spiridonov for many helpful
discussions. This work was supported in part by the Russian
Fund for Fundamental Research and the Russian  Ministry of
Education.

\end{multicols}
\end{document}